\documentstyle[aps,pre,psfig]{revtex}
\begin{document}
\title{Elastic property of  single double-stranded
DNA molecules: Theoretical study and  comparison with experiments}

\author{Zhou Haijun$^{1,2}$\cite{zhouhj},
Zhang Yang$^1$, and Ou-Yang Zhong-can$^{1,3}$}

\address{$^1$Institute of Theoretical Physics, Academia
Sinica, P.O.Box 2735, Beijing 100080, China\\
$^2$State Key Lab. of Scientific and Engineering 
Computing, Beijing 100080, China\\
$^3$Center for Advanced Study, Tsinghua University,
Beijing 100084, China}

\date{March 31, 2000, to appear in PRE}

\maketitle

\begin{abstract}

This paper aims at a comprehensive understanding on the novel 
elastic property
of double-stranded DNA (dsDNA) discovered very recently through 
single-molecule manipulation techniques. A general elastic model for
double-stranded biopolymers is proposed and a new structural parameter
called  the folding angle $\varphi$  is introduced to characterize their 
deformations. The mechanical property of long dsDNA molecules is 
then studied based on this model, where the base-stacking interactions
between DNA adjacent nucleotide basepairs, the steric effects of
basepairs,  and the electrostatic interactions along DNA backbones are
taken into account. Quantitative results are obtained by using
path integral method, and excellent agreement between theory and
the observations reported by five major experimental groups are attained.
The strong intensity of the base-stacking interactions
 ensures the structural stability of
DNA, while the short-ranged nature of such  interactions makes
externally-stimulated large structural fluctuations  possible. The
entropic elasticity, highly extensibility, and supercoiling property of
DNA are all closely related to this account. 
The present work also suggests
the possibility that  negative torque can induce structural transitions in 
highly extended DNA from right-handed B-form to left-handed configurations 
similar with  Z-form configuration. Some formulae concerned with the
application of path integral method to polymeric systems are listed in the 
Appendix. 

\end{abstract}

\pacs{87.15.By, 05.50.+q, 64.60.Cn, 75.10.Hk}

\section{Introduction}
\label{sec1}

DNA molecule is  the primary genetic material of most organisms.
It is a double-helical biopolymer in which two chains of 
complementary nucleotides (the subunits whose sequence constitutes
the genetic message) wind (usually right-handedly) around a common 
axis to form a double-helical structure \cite{saenger84}. 
Because of this unique structure, the elastic property of DNA molecule 
influences its biological functions greatly.  There are mainly three kinds
of deformations  in DNA double-helix: stretching and
bending of the molecule,  twisting of one nucleotide chain relative
to its counterpart. All these deformations have vital biological 
significance. During DNA replication, hydrogen bonds between the
complementary DNA bases should be broken and the two nucleotide chains be
separated. This strand-separation process requires cooperative unwinding
of the double-helix \cite{watson87}. 
In DNA recombination reaction, RecA proteins 
polymerize along DNA template and the DNA molecule is stretched to 
$1.5$ times its relaxed contour length \cite{nishinaka97,nishinaka98}. 
It is suspected that
thermal fluctuations of DNA central axis might be very important
for RecA polymerization \cite{leger98}.  Another important example is the 
process of chromosome condensation during prophase of the cell cycle,
 where the long (circular) DNA chain wraps tightly onto 
histone proteins and is
severely bent \cite{watson87}. 
Further more, in living cells DNA chain is usually closed,
i.e., the two ends of the molecule is linked 
together by covalent bonds and
the molecule becomes endless. With this chain-closing process,
all those quantities characterizing the topological state of the 
chain are fixed and  can   only be 
changed externally  by topoisomerases,
which are capable of transiently cutting one or both  DNA
strands and making one strand pass through the other at the cutting
point (in the case of type I topoisomerases \cite{stewart98})
 or one segment of DNA
pass through another (in the case of type II 
topoisomerases \cite{rybenkov97,yan99}). 
It is possible that this kind of enzyme-induced topology-changing
processes are also closely related to the 
particular mechanic property of DNA molecule.
For example, the frequency of collisions between two distant DNA
segments in a circular DNA molecule is influenced by the different
knot types and different linking numbers (for a definition of
this quantity, see below and Sec. \ref{sec2C}). 
 A thorough 
investigation of the  deformation and
elasticity of DNA will enable us to gain better understanding on many
important biological processes concerned with life and growth.

Detailed study on DNA elasticity now becomes possible with the
recent experimental developments, including, e.g., optical tweezer methods,
atomic force microscopy, fluorescence microscopy. These
techniques make it possible to  manipulate  directly
single polymeric molecules and to record
their elastic responses with high precision. 
Experiments done on double-stranded DNA (dsDNA) 
 have revealed that this molecule has
very novel elastic property 
\cite{smith92,bensimon95,cluzel96,smith96,strick96,strick98,allemand98,leger99}.
 When a torsionally relaxed DNA is pulled
with a force less than $10$ picoNewton (pN), its elastic response
can be quantitatively understood by regarding the chain as an
inextensible thin string with certain bending rigidity (namely,
the wormlike chain model 
\cite{smith92,bustamante94,marko95a}). However, if the external force 
is increased
up to $65$ pN, DNA chain becomes highly extensible. At this force,
the molecule transit to an over-stretched configuration termed 
S-DNA, which is $1.6$ times longer than the same molecule in its 
standard B-form structure \cite{cluzel96,smith96}.
Besides external forces, it is also possible
to apply torsional constraints to DNA double-helix by external torques.
The linking number of DNA, i.e., the total topological turns one
DNA strand winds around the other, can be fixed at a value larger
(less) than the molecule's relaxed value. In such cases we say the
DNA molecule is positively (negatively) supercoiled. It is shown 
experimentally \cite{strick96} that when external force is less than a
threshold value of about $0.3$ pN, 
the extension of DNA molecule decreases with increasing twisted stress 
and the elastic response of positively
supercoiled DNA is similar to that of negatively supercoiled DNA,
indicating the DNA chain might be regarded as achiral. However, 
if the external force is increased to be larger than this threshold,
negatively and positively supercoiled DNA molecules 
behave quite differently. Under the condition of fixed 
external force between $0.3$ pN and $3$ pN,
while positive twist stress keeps shrinking the DNA polymer,
the extension of negatively supercoiled 
DNA is insensitive to supercoiling degree \cite{strick96,strick98}. 
In higher force region, it is suggested by some authors that 
positively supercoiled DNA may transit to a configuration called 
Pauling-like DNA (P-DNA) with exposed nucleotide bases \cite{allemand98}, 
while negative torque may lead to strand-separation in DNA
molecule (denaturation of DNA double-helix \cite{strick98}). 
A very recent systematic observation
performed by L${\rm\acute{e}}$ger 
{\it et al.} \cite{leger99}, on the other hand, 
suggested another possibility that negative supercoiling may result in
left-handed Z-form configuration in DNA.

The above-mentioned complicated 
elastic property revealed by experiments may be
directly related to the versatile roles played by DNA molecule in 
living organisms. Theoretically, to understand DNA elastic property is of 
current interest. Concerned with one or another aspect of DNA elasticity,
models were proposed and valuable insights were obtained
 (see, for example, Refs.\ 
\cite{bustamante94,marko95a,hao89,fain96,marko97a,moroz97,marko97b,ha97,cizeau97,bouchiat98,nelson98,zhou98}), 
and now it is widely accepted that  the competition between DNA bending
and torsional deformations deserves to pay considerable attention 
in understanding the elastic property of DNA molecules. 
However, it is still a great challenge to understand systematically 
and quantitatively
all aspects of DNA mechanical property based on the same unified framework.
What is the intrinsic reason for DNA molecule's entropic elasticity,
highly extensibility as well as its supercoiling property? 
Is it possible for  negative torque to stabilize left-handed 
DNA configurations?  
These are just some examples of unsolved questions. 

In the present work, we have tried to obtain a comprehensive 
and quantitative understanding on  DNA mechanical property. 
We have thought that the double-stranded nature of DNA structure should be 
extremely important to its elastic property, and therefore have
constructed 
a general  elastic model in which this characteristic is properly taken into
account via the introduction of a new structural parameter, the folding
angle $\varphi$.
The elastic property of long dsDNA molecules was 
then studied based on this model, where the base-stacking interactions
between DNA adjacent nucleotide basepairs,  their  steric effects,
and  the electrostatic interactions along DNA backbones were
all considered. Quantitative results were obtained by using
path integral method, and excellent agreement between theory and
the experimental observations of several   groups were attained.
It was revealed that, on one hand,  the strong intensity of
the base-stacking  interactions 
ensures the structural stability of DNA molecule; while on the other hand,
the short-ranged nature of such interactions makes
externally-stimulated large structural fluctuations possible. The
entropic elasticity, highly extensibility, and supercoiling property of
dsDNA molecule are all closely related to this fact. 
The present work also revealed
the possibility that negative torque can induce structural transitions in 
highly extended DNA from right-handed B-form configuration
 to left-handed Z-form-like configurations. 
Some discussions on this  
respect were performed and we suggested  that a possible direct
way to check the validity of this opinion is to measure the values of
the critical torques under which such transitions are anticipated to take
place by the present calculations.

This paper is organized as follows: In  Sec.\ \ref{sec2}
 we  introduce the elastic model  for dsDNA biopolymers. 
At  the action of an external force, the elastic response
of dsDNA molecules is investigated in  Sec. \ref{sec3} and
compared  with experimental observations of 
Smith {\it et al.} \cite{smith92,smith96} and 
Cluzel {\it et al.} \cite{cluzel96}.
Section \ref{sec4} focuses on  supercoiled dsDNA
molecules, where the relationship between 
 extension and supercoiling degree  is obtained numerically and
compared with the experiment of Strick {\it et al.} \cite{strick96}. 
From the calculated folding angle distribution, we infer that
negative torque can cause structural transitions in dsDNA molecules
from  right-handed double-helix to left-handed ones.
Section \ref{sec5} is reserved for conclusion. Two appendices
are also presented: in Appendix \ref{appendixA} 
  we review some basic ideas on the 
application of path integral method in polymer physics; and
in Appendix \ref{appendixB} we list the matrix elements of the
operators in Eq. (\ref{eq13}) and Eq. (\ref{eq18}).
Some parts of this work
 have been briefly reported  in a previous letter\cite{zhou992}.


\section{Elastic model of double-stranded DNA molecule}
\label{sec2}

As already stressed, DNA molecule is a double-stranded biopolymer. 
Its two complementary sugar-phosphate chains twist
around each other to form a right-handed double-helix. Each chain is 
a linear polynucleotide consisting of the following  four bases: two purines 
(A, G) and two pyrimidines (C, T) \cite{saenger84,watson87}. 
The two chains are 
joined together by  hydrogen bonds between pairs of nucleotides
A-T and G-C. Hereafter, we refer to the two sugar-phosphate chains as 
the {\it backbones} and the hydrogen-bonded pairs of nucleotides as 
the {\it basepairs}. 
 In this section  we discuss the energetics 
of such an elastic system (see Fig.\ \ref{fig1}). First of all, 
the bending energy of the  backbones of
such double-stranded polymers will be considered; then we will
discuss the interactions between DNA basepairs and energy terms related with
external fields.

\subsection{Bending and folding deformations}
\label{sec2A}

The backbones can be regarded as two inextensible
wormlike chains characterized with a very small bending rigidity 
$\kappa=k_B T\ell_p$, where $k_B$ is Boltzmann's constant and $T$
the environmental temperature, and 
$\ell_p\simeq 1.5$ nm is the bending persistence 
length of single-stranded DNA (ssDNA) chains \cite{smith96}. 
The bending energy of 
each backbone is thus expressed as $\kappa \int_0^L(d{\bf t}_i/d s)^2
d s$, where ${\bf t}_i(s)$ $(i=1, 2)$ is the unit tangent
vector at arclength $s$ along the $i$-th backbone 
\cite{everaers95,liverpool98}, and $L$ is the total contour
length of each backbone. The
position vectors of the two backbones are expressed as 
${\bf r}_i(s)=\int^s {\bf t}_i (s^\prime) d s^\prime$. 

\begin{figure}[t]
\vskip -4.5cm
\hspace{4.2cm}\psfig{figure=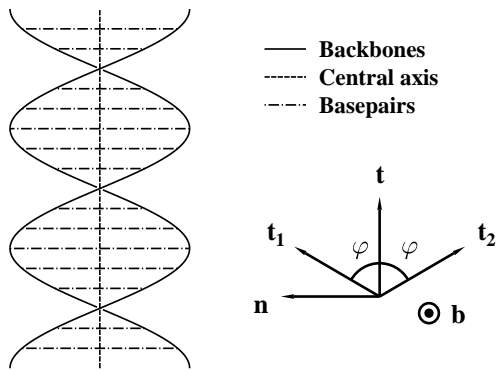,height=12.0cm}
\vskip -1.5cm
\caption{Schematic representation  of a double-stranded DNA model 
used in this paper. The right part demonstrates the
 definition of \protect\( \varphi\protect\) on 
 the local \protect\({\bf t}-{\bf n}\protect\) plane, 
where \protect\({\bf t}\protect\), \protect\({\bf t}_1\protect\) and
\protect\({\bf t}_2\protect\) are, respectively, the tangential 
vectors of the central
axis and the two backbones; \protect\(\varphi\protect\) 
is the folding angle; and the 
unit vector \protect\({\bf b}={\bf t}\times {\bf n}\protect\) is perpendicular
to the \protect\({\bf t}-{\bf n}\protect\) plane. }
\label{fig1}
\end{figure}

Since there are many relatively rigid
basepairs between the two backbones
in many cases the lateral distance between the backbones can
be regarded to be  constant and equal to $2R$.\footnote{In 
our present work, we have not taken into account the possible
deformations of the nucleotide basepairs. In many cases
this may be a reasonable assumption. However, 
under some extreme conditions such kind of deformations 
may turn to be important. For example, when DNA double-helix
is stretched and at the same time is 
applied with a large positive torque, the nucleotide basepairs
may collapse (see Ref. \cite{allemand98} for a description).}
In this subsection we focus on the bending energy of the backbones,
therefore, for the moment we regard each basepair as a rigid rod
of length $2R$ linking 
between the two backbones and pointing along direction 
denoted  by a unit vector ${\bf b}$ 
from ${\bf r}_1$ to ${\bf r}_2$ (Fig. \ref{fig1}).
Then, ${\bf r}_2(s)-{\bf r}_1(s)=2 R {\bf b}(s)$. 
In B-form DNA the basepair plane is perpendicular to DNA
axis, therefore, in  our model 
relative sliding of the two backbones is
not considered and  the basepair rod is thought to be
perpendicular to both backbones \cite{liverpool98}, 
with ${\bf b}(s)\cdot {\bf t}_1 (s)
={\bf b}(s)\cdot {\bf t}_2 (s)\equiv 0$. The central axis of the
double-stranded polymer can be defined as ${\bf r}(s)=
{\bf r}_1(s)+R {\bf b}(s)$ $[= {\bf r}_2 (s)-R {\bf b}(s) =
({\bf r}_1 (s)+{\bf r}_2 (s))/2]$, and its tangent vector is
denoted by ${\bf t}$. In consistence with
 actual DNA structures, the central axial tangent ${\bf t}$
is also perpendicular to ${\bf b}$, i.e., ${\bf b}(s)\cdot {\bf t}(s)
=0$. (Notice that, however,  ${\bf t}(s)\neq d{\bf r}/ds$; in 
this paper, $s$ always refers to the arclength of the {\it backbones}.)

Since all the tangent vectors ${\bf t}_1$, ${\bf t}_2$, and 
${\bf t}$ lie on the same plane perpendicular to ${\bf b}$, we 
can write that 
\begin{equation}
\label{eq1}
\begin{array}{l}
{\bf t}_1(s)={\bf t}(s)\cos\varphi (s)+ {\bf n}(s)\sin\varphi (s), \\
{\bf t}_2(s)={\bf t}(s)\cos\varphi (s)-{\bf n}(s)\sin\varphi (s),
\end{array}
\end{equation}
where ${\bf n}$ is also a unit vector and 
${\bf n}={\bf b}\times {\bf t}$, and 
$\varphi$ is defined as half the rotational angle from ${\bf t}_2$
to ${\bf t}_1$, with ${\bf b}$ being the rotational axis (Fig.\ \ref{fig1}).
We call $\varphi$ the {\it folding angle}, and it can vary in the 
range $(-\pi/2,\;\;+\pi/2)$, with $\varphi >0$ corresponding to
right-handed rotations and hence right-handed double-helical
configurations and $\varphi <0$ to left-handed ones. With the
help of Eq. (\ref{eq1}), we know that 
\begin{equation}
\label{eq2}
{d {\bf b}\over d s}={{\bf t}_2 -{\bf t}_1 \over 2 R}=- {\sin\varphi
\over R} {\bf n},
\end{equation}
and 
\begin{equation}
\label{eq3}
{d{\bf r}\over ds }={1\over 2}({\bf t}_1+{\bf t}_2)={\bf t}\cos\varphi.
\end{equation}
Equation (\ref{eq3}) indicates that $\cos\varphi$ measures the
extent to which the backbones are $``$folded" with respect to the
central axis. Based on  Eqs. (\ref{eq1}-\ref{eq3}), the total bending
energy of the  two backbones can be expressed in the following 
form:
\begin{eqnarray}
\label{eq4}
E_b &  =  & {\kappa\over 2}\int_0^L ({d{\bf t}_1\over ds})^2 ds +
{\kappa\over 2}\int_0^L ({d{\bf t}_2\over ds})^2 ds \nonumber \\
 &  =  & \kappa \int_0^L ds \left[({d{\bf t}\over ds})^2\cos^2\varphi
+\sin^2\varphi ({d{\bf n}\over ds})^2+({d\varphi\over ds})^2\right] \nonumber \\
 & = & \int_0^L ds \left[\kappa ({d{\bf t}\over d s})^2+
\kappa ({d\varphi\over d s})^2+{\kappa\over R^2}\sin^4\varphi\right].
\end{eqnarray}
The bending energy is thus decomposed into the bending energy of
the central axis (the first term of Eq. (\ref{eq4})) plus 
the {\it folding energy} of the backbones (the second and 
third terms of Eq. (\ref{eq4})). The physical meanings of
these two energy contributions are very clear, and Eq. (\ref{eq4})
is very helpful for our following calculations. In Eq. (\ref{eq4}), the
bending energy of the central axis is very similar with that of a 
wormlike chain \cite{marko95a}, 
both of which are related to the  square of the  changing rate
of the axial tangent vectors.  But there are two  important differences:
(a) in the derivative $d{\bf t}/ds$ of Eq. (\ref{eq4}),
 the arclength parameter
$s$ is measured along the backbone,  not along the central axis; 
(b) in the wormlike chain model the central axis is inextensible, while
 here the central axis is extensible.   

In deriving Eq. (\ref{eq4}), the basepairs are models just as thin
rigid rods of fixed length, i.e., the 
DNA molecule is viewed as a  ladder-like structure (see Fig.\ \ref{fig1}). 
Actually, however, basepairs form disc-like
structures and have finite volume. The steric effects caused by 
the finite volume of basepairs is anticipated to hinder considerably
the bending deformation of the central axis, and hence will increases 
its bending rigidity greatly \cite{saenger84}. 
 Furthermore, dsDNA molecule is a strong
polyelectrolyte, with  negatively-charged groups distributed 
regularly along the chain's surface. 
The electrostatic repulsion force between
these negatively-charged groups will also considerably increase the
bending rigidity of the dsDNA chain \cite{saenger84}.  To quantitatively take
 into account the above mentioned two kinds of effects is very difficult.
Here we treat this problem phenomenologically by
simply  replacing  the bending rigidity
$\kappa$ in the first term of Eq. (\ref{eq4}) with a 
quantity $\kappa^*$.
It is required that $\kappa^* > \kappa$, 
and the precise value of $\kappa^*$ will then
be determined self-consistently 
by the best fitting with experimental data as shown in Sec.\ \ref{sec3}.

\subsection{Base-stacking interactions between basepairs}
\label{sec2B}

In Sec. {\ref{sec2A}, we have discussed in detail the bending energy of
dsDNA polymers, which is caused by bending of the backbones as well as 
steric effects  and electrostatic interactions. In dsDNA 
molecule there is another kind of important interactions, namely 
the base-stacking interaction between adjacent nucleotide basepairs 
\cite{saenger84,watson87}.
The base-stacking interactions originate from the weak
van der Waals attraction  between the polar groups in 
adjacent nucleotide basepairs. Such interactions are short-ranged
and their total effect is usually described by a potential energy of
the Lennard-Jones form ($6$-$12$ potential \cite{saenger84}). 
Base-stacking interactions play significant role in 
stabilization of DNA double-helix.
The main reason why  DNA can but RNA  can not form long double-helix is 
as follows\cite{stryer}: 
Because of the  steric interference caused by the hydroxyl 
group attached to  the $2^\prime$
carbon of RNA riboses, the stacking interaction between adjacent  RNA 
nucleotide basepairs is very weak and can not stabilize the formed 
double-helical structure;  while in the DNA ribose, it is a hydrogen atom
attached to its $2^\prime$ carbon and  serious steric 
interference is avoided (fortunately!).

In a continuum theory of elasticity,  the summed  total base-stacking
potential energy is converted into the form of the following integration:
\begin{equation}
\label{eq5}
E_{LJ}=\sum\limits_{i=1}^{N-1} U_{i,i+1}=\int_0^L \rho(\varphi) d s,
\end{equation}
where $U_{i,i+1}$ is the  base-stacking potential  between the $i$-th
and the $(i+1)$-th basepair, $N$ is the total number of basepairs, and 
the base-stacking energy density $\rho$ is expressed as 
\begin{equation}
\label{eq6}
\rho(\varphi)=\left\{
\begin{array}{lcll}
{\epsilon\over r_0} [({\cos\varphi_0\over \cos\varphi})^{12} &-& 
2({\cos\varphi_0\over \cos\varphi})^6] & \;\;\;\;({\rm for}\;\;
\varphi\geq 0), \\
{\epsilon\over r_0}[\cos^{12}\varphi_0 &-& 2 \cos^6\varphi_0]
& \;\;\;\;({\rm for}\;\; \varphi <0).
\end{array}
\right.
\end{equation}
In Eq. (\ref{eq6}), the parameter $r_0$ is the
 {\it backbone} arclength between 
adjacent bases ($r_0=L/N$); $\varphi_0$ is a parameter related to the 
equilibrium distance between a DNA dimer ($r_0\cos\varphi_0 \sim 
3.4$ $\AA$); and $\epsilon$ is the
base-stacking intensity which is
generally base-sequence specific \cite{saenger84}. 
In this paper we focus on
macroscopic properties of long DNA chains composed of relatively 
random sequences,  therefore  we 
just consider $\epsilon$ in the average
sense and take it as a constant, with $\epsilon$ $\simeq
14.0$ $k_B T$ as averaged over quantum-mechanically  calculated
results on all the different DNA dimers \cite{saenger84}. 

The asymmetric base-stacking
potential Eq. (\ref{eq6}) ensures a relaxed DNA to take on a 
right-handed double-helix configuration (i.e., the B-form) 
with its folding angle
$\varphi \sim \varphi_0$.
To deviate the local configuration of DNA considerably from
its B-form  generally requires a free energy 
of the order of  $\epsilon$ per
basepair. Thus, DNA molecule will be very stable under normal
physiological conditions and thermal energy can only make it 
fluctuate  very slightly around its equilibrium configuration, since
$\epsilon \gg kT$.
Nevertheless,  although the stacking intensity 
$\epsilon$ in dsDNA is very strong compared with 
thermal energy, the base-stacking interaction by its nature is
short-ranged  and hence sensitive 
to the distance between the adjacent basepairs. If dsDNA chain is
stretched by large  external forces, which cause  the average inter-basepair
distance to exceed some threshold value determined intrinsically by the
molecule, the restoring force provided by the base-stacking interactions
will no longer be able to offset the external forces. 
Consequently, it will be possible that  the B-form configuration of 
dsDNA will collapse and the chain will turn to be  highly extensible.
Thus, on one hand, the strong base-stacking interaction ensures
 the standard  B-form configuration to be very
stable upon thermal fluctuations and small external forces (this
is required for the biological functions of DNA molecule to be 
properly fulfilled \cite{watson87});
but on the other hand, its short-rangedness gives it
considerable latitude to change its configuration to adapt to 
possible  severe environments (otherwise, the chain may be 
pulled break by external forces, for example, during DNA segregation 
\cite{watson87}). 
This property of DNA base-stacking interactions is very important
to dsDNA molecule. As we will see in Secs.\ \ref{sec3} and
\ref{sec4}, the  mechanical property of DNA chain is indeed
 closely related to the above-mentioned
insight. 

\subsection{External forces and torques}
\label{sec2C}

In the previous two subsections, we have described
 the intrinsic  energy of DNA double-helix. 
Experimentally, to probe the elastic response of  linear
DNA molecule,  the polymer chain is  often 
pulled  by external force fields  and/or
  untwisted or overtwisted by
external  torques.
To study  the mechanic response  of
dsDNA molecule, we consider in this subsection
the energy terms related to 
external forces and  torques in our theoretical framework. 

For the external force fields, here we constrained ourselves to
the simplest situation where one  terminal of DNA molecule is fixed and
the other terminal is pulled with a force ${\bf F}=
f {\bf z}_0$ along direction of unit vector ${\bf z}_0$ \cite{smith92}. 
(In fact, hydrodynamic fields or electric fields are also frequently used to
stretch semiflexible polymers \cite{perkins97}, but we will 
not discuss such cases in this paper.)  
The end-to-end vector
of a DNA chain is expressed as $\int_0^L {\bf t}(s)\cos\varphi(s) d s$
according to Eq. (\ref{eq3}).
Then the total $``$potential" energy of the chain in the external force
field is
\begin{equation}
E_f=-\int_0^L {\bf t} \cos\varphi ds \cdot {\bf F}= -\int_0^L f
{\bf t}\cdot {\bf z}_0 \cos\varphi d s.
\label{eq7}
\end{equation}

In the experimental setup, external torques can be applied on linear
dsDNA molecule by the following procedure: first,  DNA ligases are used to
ligate all the possible single-stranded nicks; then the two
strands of dsDNA molecule at one end are fixed onto a template, 
while the two strands at the 
other end is attached tightly to a magnetic bead; afterwards,
 torques are  introduced into DNA double-helix by 
rotating the magnet bead with an  external magnetic field 
\cite{cluzel96,allemand98}. 
Torque energy is  then related to the topological turns caused by
the external torque on DNA double-helix. The total number of  topological 
turns one DNA strand winds around the other, which is usually 
termed the total linking number $Lk$ \cite{white69,fuller71,crick76}, is
expressed as the sum of the twisting number, $Tw({\bf r}_1 ({\bf r}_2),
{\bf r})$ of backbone ${\bf r}_1$ (or ${\bf r}_2$) around the central
axis ${\bf r}$ and the writhing number
 $Wr({\bf r})$ of the central axis;
i.e., $Lk=Tw+Wr$. According to Refs. \cite{fuller71,crick76,white88} 
and Eq. (\ref{eq2}), we
obtain that
\begin{equation}
\label{eq8}
Tw({\bf r}_1, {\bf r})=
{1\over 2\pi}\int_0^L {\bf t}\times {\bf b}\cdot {d {\bf b}
\over d s} ds ={1\over 2\pi}\int_0^L {\sin\varphi \over R} ds.
\end{equation}
The writhing number of the central axis is generally much more 
difficult to calculate.  It is expressed as the following Gauss 
integral over the central axis \cite{white69}:
\begin{equation}
\label{writhe}
Wr({\bf r})={1\over 4\pi}\int\int {{ d{\bf r}\times d{\bf r}^\prime \cdot
({\bf r}-{\bf r}^\prime)}\over |{\bf r}-{\bf r}^\prime|^3}.
\end{equation}
In the case of  linear chains, 
provided that some fixed direction (for example 
the  direction of the external force, ${\bf z}_0$) can
be specified and that the tangent vector ${\bf t}$ never
points to $-{\bf z}_0$ (i.e., 
 ${\bf t}\cdot {\bf z}_0\neq -1$), it was proved by  Fuller
 that the writhing number Eq. (\ref{writhe}) 
can be calculated alternatively according to the following
formula \cite{fuller78}:
\begin{equation}
\label{eq9}
Wr({\bf r})={1\over 2\pi}\int\limits_0^L {{{\bf z}_0\times {\bf t} 
\cdot d{\bf t}/ d s} \over {1+{\bf z}_0\cdot {\bf t}}} d s.
\end{equation}
The above equation can be further simplified for
 highly extended linear DNA chains whose tangent ${\bf t}$
fluctuates only slightly around ${\bf z}_0$. In this case,
 Eq. (\ref{eq9}) leads
to the  approximate expression that
\begin{equation}
\label{eq10}
Wr({\bf r})\simeq {1\over 4\pi}\int_0^L \left[-t_y {d t_x \over
d s}+t_x {d t_y \over ds}\right] ds,
\end{equation}
where $t_x$ and $t_y$ are, respectively, the two components of ${\bf t}$
with respect to two arbitrarily chosen  orthonormal  directions 
(${\bf x}_0$ and ${\bf y}_0$) on the plane perpendicular to ${\bf z}_0$.

The energy caused by the  external torque of magnitude $\Gamma$ is then 
equal to 
\begin{equation}
\label{eq11}
E_t=-2\pi\Gamma Lk=-2\pi\Gamma (Tw+Wr).
\end{equation}

To conclude  this section, the total energy of a dsDNA molecule under
the action of an external force and an external torque is  expressed as
\begin{eqnarray}
E & = & E_b+E_{LJ}+E_f+E_t \nonumber \\
& = & \int_0^L\left[ \kappa^* ({d {\bf t}\over ds})^2+\kappa ({d\varphi
\over ds})^2+{\kappa \over R^2}\sin^4\varphi+\rho(\varphi)-f{\bf t}\cdot
{\bf z}_0\cos\varphi-{\Gamma\over R}\sin\varphi-\Gamma {{\bf z}_0\times
{\bf t}\cdot d{\bf t}/ds \over 1+{\bf z}_0\cdot {\bf t}}
\right] ds \label{eq12a} \\
 & \simeq & \int_0^L\left[ \kappa^* ({d {\bf t}\over ds})^2+\kappa ({d\varphi
\over ds})^2+{\kappa \over R^2}\sin^4\varphi+\rho(\varphi)-f{\bf t}\cdot
{\bf z}_0\cos\varphi-{\Gamma\over R}\sin\varphi+{\Gamma\over 2}
t_y{d t_x\over ds}-{\Gamma\over 2} t_x {d t_y\over ds}\right] ds. \label{eq12}
\end{eqnarray}
Notice that Eq. (\ref{eq12}) can be applied only in the case of
highly extended DNA.  In the following two  sections, we will study the 
mechanical property of single dsDNA molecules based on the
model energy  Eq. (\ref{eq12a}) and Eq. (\ref{eq12}). The theoretical results
will be compared with experimental observations and discussed. 

\section{Extensibility and entropic elasticity of DNA}
\label{sec3}

In this section we investigate the elastic responses of single 
DNA molecules
under the actions of external forces based on 
the model introduced in  Sec.\ \ref{sec2}.
There is no external torque acted, thus $\Gamma=0$ in
Eq. (\ref{eq12a}). The particular form of the energy function 
Eq. (\ref{eq12a}) of the present model makes it
convenient for us to study its statistical property
by path integral method. In Appendix \ref{appendixA} a detailed
description on the application of path integral method to polymer 
physics is given \cite{zhounote1}. Our 
calculations in this and the next sections  are based on this method.

For a polymer whose energy is expressed in the form of Eq. (\ref{eq12a})
with $\Gamma=0$, according to the technique outlined in
Appendix \ref{appendixA} (see Eqs. (\ref{eqA1}) and (\ref{eqA7})), 
the Green equation governing the evolution of the
$``$wave function" $\Psi({\bf t}, \varphi; s)$ of the 
 system is obtained to be of the 
following form:
\begin{equation}
{\partial \Psi({\bf t},  \varphi; s)\over \partial s}=
\left[{\partial^2 \over 4\ell_p^*\partial 
{\bf t}^2}+{\partial ^2\over 4\ell_p\partial\varphi^2}+{f
\cos\varphi\over k_B T} {\bf t}\cdot {\bf z}_0-{\rho(\varphi)\over k_B T}
-{\ell_p\over R^2}\sin^4\varphi \right] \Psi({\bf t}, \varphi; s),
\label{eq13}
\end{equation}
where $\ell_p^*=\kappa^*/k_BT$ and $\ell=\kappa/k_B T$.
The spectrum of the above Green  equation is discrete and, 
for a long dsDNA molecule according to Eqs. (\ref{eqA10}) and
(\ref{eqA13}),   
its average extension can be
obtained either by differentiation of
the ground-state eigenvalue, $g_0$, of Eq. (\ref{eq13}) 
with respect to $f$:
\begin{equation}
\label{eq14}
\langle Z\rangle=\int_0^L\langle {\bf t}\cdot {\bf z}_0 \cos\varphi
\rangle d s=-L k_B T {\partial g_0\over \partial f},
\end{equation}
or by a direct integration with the normalized 
ground-state eigenfunction,
$\Phi_0({\bf t},\varphi)$, of Eq. (\ref{eq13}):
\begin{equation}
\label{eq15}
\langle Z\rangle=L\int|\Phi_0 |^2 {\bf t}\cdot {\bf z}_0 
\cos \varphi d {\bf t}
d \varphi.
\end{equation}
Both $g_0$ an $\Phi_0({\bf t},\varphi)$ can be obtained 
numerically through standard 
diagonalization methods 
 and identical results 
are obtained by 
Eqs. (\ref{eq14}) and (\ref{eq15}).
Here we just briefly outline the main procedures in 
converting Eq. (\ref{eq13}) into the form of a matrix.

\begin{figure}[t]
\vskip 3.5cm
\hspace{1.2cm}\psfig{figure=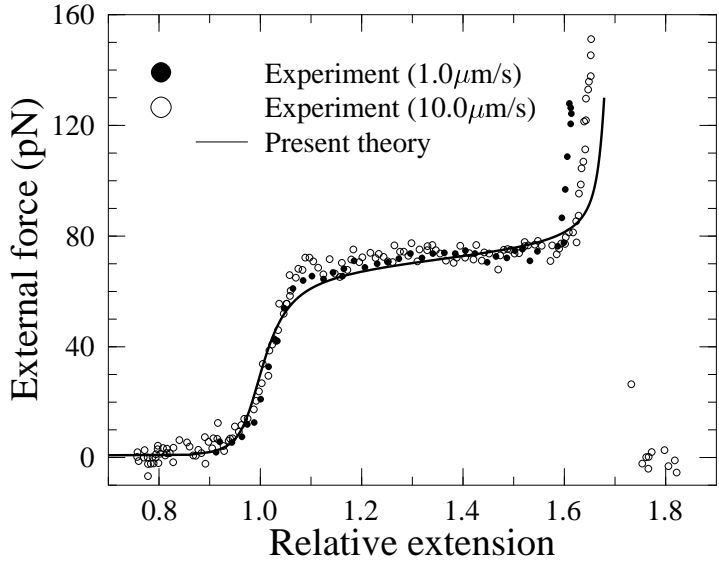,height=9.0cm}
\vskip -2.5cm
\caption{
Force-extension relation of torsionally relaxed  DNA molecule.
Experimental data  is from
Fig. 2A of {\protect\cite{cluzel96}}(symbols). Theoretical curve is
obtained by the following considerations: (a) 
\protect\(\ell_p=1.5\protect\) nm 
 and \protect\(\epsilon=14.0 k_B T\protect\);
 (b) \protect\(\ell_p^*=53.0/2\langle\cos\varphi\rangle_{f=0}\protect\) nm, 
\protect\(r_0=0.34/\langle\cos\varphi\rangle_{f=0}\protect\) nm and
\protect\(R=(0.34\times 10.5/2\pi)\langle \tan\varphi\rangle_{f=0}\protect\)
nm;
(c) adjust the value of \protect\(\varphi_0\protect\) to fit the data.
For each \protect\(\varphi_0\protect\), the value of 
\protect\(\langle\cos\varphi\rangle_{f=0}\protect\) is 
obtained self-consistently. 
The present curve is drawn with
\protect\(\varphi_0=62.0^{\circ}\protect\) (in close consistence with the
 structural property of DNA),  
 and \protect\(\langle\cos\varphi\rangle_{f=0}\protect\) is
determined to be \protect\(0.573840\protect\). DNA extension is scaled with
 its {\protect {\bf B}}-form contour length
\protect\(L \langle\cos\varphi\rangle_{f=0}\protect\).
}
\label{fig2}
\end{figure}

Firstly, for our convenience we perform the following
transformation:
\begin{equation}
\varphi=\tilde{\varphi}-{\pi\over 2},
\label{varphi}
\end{equation}
hence the new argument $\tilde{\varphi}$ can change in the range from
$0$ to $\pi$. Then,
 we choose the combination of 
$Y_{lm}({\bf t})$ and $f_n(\tilde{\varphi})$ as the
base functions of the Green equation Eq. (\ref{eq13}):
\begin{equation}
\label{base}
\Psi({\bf t}, \tilde{\varphi}; s)=\sum_{lmn} C_{lmn}(s) Y_{lm}({\bf t})
f_n(\tilde{\varphi}).
\end{equation}

\noindent
In the above expression, $Y_{lm}({\bf t})=Y_{lm}(\theta,\phi)$ 
($l=0, 1, 2, \cdots; m=0, \pm 1,  \cdots, \pm l)$
are the  spherical harmonics \cite{zeng99}, where $\theta$ and
$\phi$ are the two directional angles of ${\bf t}$, i.e., ${\bf t}=
(\sin\theta\cos\phi, \sin\theta\sin\phi, \cos\theta)$;
and 
\begin{equation}
\label{fn}
f_n(\tilde{\varphi})=
\sqrt{2 \over a}\sin({n\pi \over a}\tilde{\varphi}) \;\;\;\;(n=1, 2, \cdots)
\end{equation}
are  the eigenfunctions of one-dimensional 
infinitely deep square potential well of width $a$.
\footnote{In the actual calculations, the right 
boundary of the square well is chosen to be slightly
less than $\pi$ to avoid  flush-off of computer
memory caused by the divergence of the base-stacking
potential Eq. (\ref{eq6}) at $\tilde{\varphi}=\pi$. 
We set $a=0.95\pi$ in this paper. However, we have
checked that the results are almost identical for other values of
$a$, provided that $a\geq 165^\circ$.}  

\begin{figure}[t]
\vskip 3.5cm
\hspace{1.2cm}\psfig{figure=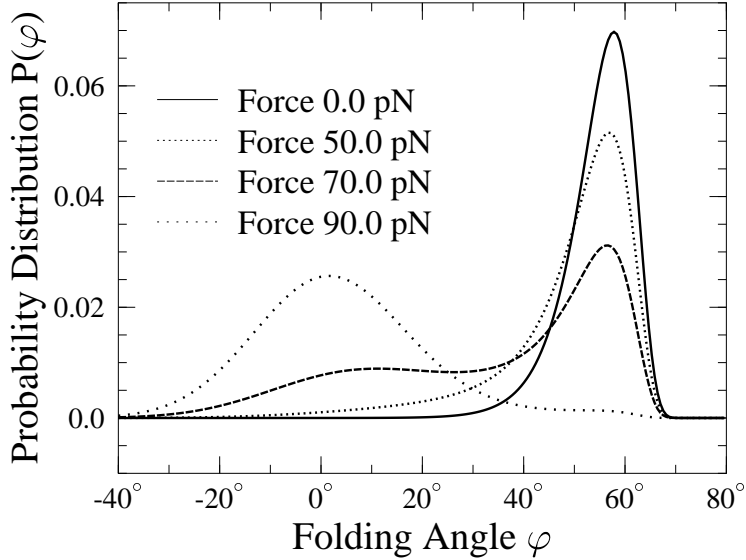,height=9.0cm}
\vskip -2.5cm
\caption{Folding angle distribution for torsionally relaxed
DNA molecules under external forces.}
\label{fig3}
\end{figure}

With the wave function $\Psi$
being expanded using the above mentioned
base functions,  the operator acting on 
$\Psi$ (i.e., the expression in the 
square brackets of Eq. (\ref{eq13})) can
also be written into matrix form under these base functions.
This matrix, whose
elements are listed in Appendix \ref{appendixB}, 
 is then diagolized numerically to
obtain its ground-state eigenvalue and eigenfunction. 
To simplify the calculation, we further notice that in the
present case of Eq. (\ref{eq13}), the ground-state is
independent to $\phi$, i.e., $m$ can be set to $m=0$ in
Eq. (\ref{base}).

The resulting force vs extension relation 
 obtained from Eq. (\ref{eq14}) or
Eq.  (\ref{eq15}) is  shown in Fig. \ref{fig2} 
 in the whole 
relevant force range
 and compared with the experimental 
observation of Cluzel {\it
et al.} \cite{cluzel96,smith96}.
The theoretical curve in this figure is 
obtained with just one adjustable parameter (see caption of 
Fig. \ref{fig2});
the agreement with experiment is  excellent. 
Figure \ref{fig2}
demonstrates  that the highly extensibility of DNA molecule 
under large external forces can be quantitatively 
explained by the present model.

To further understand the force-induced extensibility of DNA, in
Fig. \ref{fig3}  the folding angle distribution
of dsDNA molecule is shown, with 
the external force kept at  different  values. 
Here,  according to Eq. (\ref{eqA13}) the folding angle distribution 
$P(\varphi)$ is calculated by the following formula:
\begin{equation}
P(\varphi)=\int |\Phi_0({\bf t}, \varphi)|^2 d {\bf t}.
\label{eq16}
\end{equation}

Figures \ref{fig2} and \ref{fig3}, taking together, 
demonstrate that the  elastic behaviors of  dsDNA molecule
are radically different under the condition of
 low and large applied forces. 
In the following, we will discuss them separately. 

\begin{figure}[t]
\vskip 3.5cm
\hspace{1.2cm}\psfig{figure=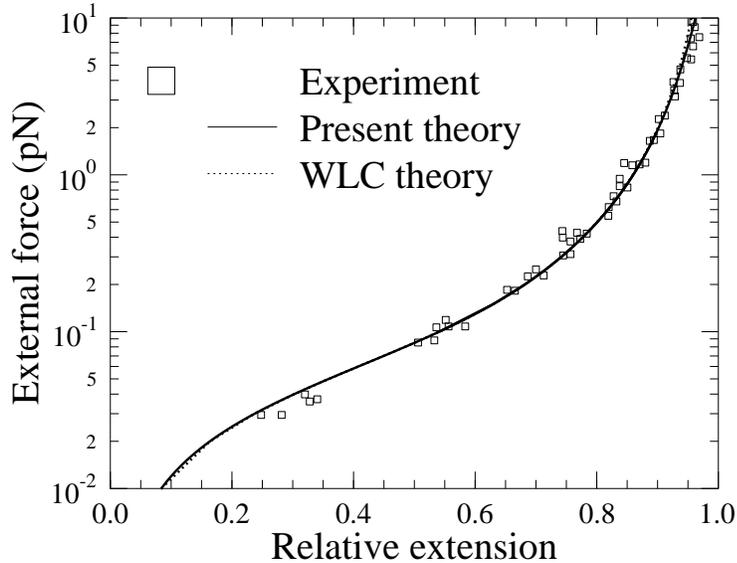,height=9.0cm}
\vskip -2.5cm
\caption{Low-force  elastic behavior of DNA. Here
experimental data is from Fig. 5B of {\protect \cite{smith92}}, 
the dotted curve is obtained for a wormlike chain with
bending persistence length  \protect\( 53.0\protect\)
 nm and the parameters for the 
slid curve are the same as those in Fig.~{\protect\ref{fig2}}.}
\label{fig4}
\end{figure}

{\it The  low-force region}\hspace{0.5cm} 
When external force is low ($\leq 10$ pN), 
the folding angle is distributed
narrowly around the angle of $\varphi\simeq + 57^\circ$, and there is no
probability for the folding angle to take on values less than $0^\circ$ 
(Fig. \ref{fig3}), indicating that 
DNA chain is completely in the right-handed B-form 
configuration with
small axial fluctuations.
This should be attributed to the strong base-stacking intensity,
as pointed out in Sec. \ref{sec2B}.
Consequently, the elasticity of DNA is solely caused by thermal 
fluctuations in the axial tangent ${\bf t}$ (Fig. \ref{fig2}), 
and DNA molecule can be regarded as 
an inextensible chain. This is the physical 
reason why, in this force region, the elastic behavior of DNA 
can be well described by the wormlike chain model 
\cite{bustamante94,marko95a,kroy96}. Indeed, as shown in
 Fig. \ref{fig4}, at forces $\leq 10$ pN, 
the wormlike chain model and the present 
model give identical results. Thus, we can conclude with confidence that,
when external fields are not strong, 
the wormlike chain model is a good approximation of the present
model to describe the elastic property of dsDNA molecules; and
the bending persistence length of the molecule is 
$ 2\ell_p^* \langle \cos\varphi\rangle$, as indicated by 
Eq. (\ref{eq3}) and Eq. (\ref{eq4}).

{\it The  large-force region}\hspace{0.5cm} 
 With the continuous increase of external pulling forces,
the axial fluctuations becomes more and more significant. 
For example,  at forces
$\simeq 50$ pN, although the folding angle distribution is still peaked
at $\varphi\simeq 57^\circ$, there is also considerable probability for
the folding angle to be distributed in the region $\varphi\sim 0^\circ$
(Fig. \ref{fig4}). Therefore, at this force region, DNA polymer
can no longer be regarded as inextensible.
 At $f\simeq 65$ pN, another peak in the folding angle
distribution begins to emerge   at $\varphi\simeq 0^\circ$, marking the onset of
cooperative transition from B-form DNA to overstretched S-form
DNA \cite{cluzel96,smith96}. 
This is closely related to the short-ranged nature of the 
base-stacking interactions \cite{saenger84} (see Sec. \ref{sec2B}). 
At even higher forces ($f\geq 80$ pN)
\footnote{This threshold $f_t$ of over-stretch force is also consistent with 
a plain evaluation from base-stacking potential of $\epsilon\sim
f_tr_0$, i.e. $f_t\sim 90$ pN.}, 
the DNA molecule becomes completely into the 
overstretched form with its folding angle
peaked at $\varphi =0^\circ$. 

The force-induced axial 
fluctuations in DNA double-helix 
can be biologically significant. 
For example, it has been demonstrated that axial  fluctuations in dsDNA
enhance considerably the polymerization of RecA proteins 
along  DNA chain  \cite{leger98,shivashankar99,hegner99}.
An quantitative study on the coupling 
between RecA polymerization and DNA axial fluctuation is
anticipated to be helpful.

It seems that in the experiments
\cite{cluzel96,smith96}  the transition to S-DNA occurs even 
more cooperatively and
abruptly than predicted by the present theory (see Fig. \ref{fig2}).
This may be related to the existence 
of single-stranded breaks (nicks) in the 
dsDNA molecules used in the   experiments.
Nicks in DNA backbones 
can  lead to strand-separation or relative 
sliding of backbones \cite{cluzel96,smith96}, 
and they  can make the transition process more cooperative. 
However, the comprehensive agreement achieved in 
Figs. \ref{fig2} and \ref{fig4} indicates that 
such effects are only of limited significance.  
The elasticity of DNA is mainly determined by the 
competition between folding angle fluctuation and tangential 
fluctuation, which are
governed, respectively, by the base-stacking 
interactions ($\epsilon$) and the axial bending rigidity ($\kappa^*$) in
Eq. (\ref{eq12a}).

\section{Elastic Property of  supercoiled DNA}
\label{sec4}

In the preceding section,  we have discussed  the 
elastic response of long DNA chains under the action
of external forces. 
In the present  section, we  turn to  study the 
elasticity of   supercoiled
DNA double-helix. For this purpose, 
in the experimental setup, all the possible nicks in the
DNA nucleotide strands are ligated \cite{strick96}, 
and a  torque as well as an external pulling force
is acting on one terminal of the DNA double-helix, 
which typically untwists or overtwists the
original B-form double-helix to some extent 
and makes its total linking number (refer to  Sec. \ref{sec2C} for 
the definition of the linking number)
less or greater than the equilibrium value. We say that
 such DNA molecules with deficit (excess)
linking number are negatively (positively) supercoiled, 
and define the degree of supercoiling as
\begin{equation}
\label{eq17}
\sigma={Lk-Lk_0\over Lk_0},
\end{equation}
where $Lk_0$ represents 
the linking number of a relaxed DNA of the same contour length.
In living organisms,  DNA molecules are
often negatively supercoiled, with a linking number deficit of
about $\sigma=-0.06$.  Thus, a detailed investigation on the
mechanical property of  supercoiled DNA molecules
is not only of academic interest but can also 
help us  to understand  the possible biological
advantages of negative supercoiling.

\subsection{Relationship between extension and supercoiling degree}
\label{sec4A}

We focus on the property of  highly extended DNA molecules
whose tangent vectors fluctuate only slightly around the force direction
${\bf z}_0$. According to what we have mentioned in Sec.\ \ref{sec2C},
in this case the approximate energy expression
Eq. (\ref{eq12}) can be used. 
 The external stretching force is restricted to
be greater than $0.3$ pN  to make sure
that the end-to-end distance of DNA chain approaches its
contour length (see Fig.\ \ref{fig5}).
Based on  Eqs. (\ref{eq12}) and (\ref{eqA7}),
 the Green equation for
highly stretched and  supercoiled dsDNA is  then
obtained to be 
\begin{eqnarray}
{\partial \Psi({\bf t}, \varphi; s)\over \partial s}
& = & \left[{\partial^2 \over 4\ell_p^*\partial {\bf t}^2}+
{\partial^2\over 4\ell_p \partial \varphi^2}+
{f\cos\varphi\over k_B T}{\bf t}\cdot {\bf z}_0 -
{\rho(\varphi)\over k_B T} -{\ell_p \over R^2}\sin^4\varphi \right. \nonumber \\
& \; & \left. +{\Gamma\over R k_B T}\sin\varphi-{\Gamma\over 4k_B T \ell_p^*}{\partial\over \partial
\phi}+{\Gamma^2\over 16 \ell_p^* (k_B T)^2}\sin^2\theta\right]
\Psi({\bf t},\varphi;s), 
\label{eq18} 
\end{eqnarray}
where $(\theta, \phi)$ are the two directional angles 
of ${\bf t}$ as mentioned before in
Sec.\ \ref{sec3}.
Similar to what we have done in Sec. \ref{sec3}, we can now
express  the above Green equation  in matrix form using the combinations of 
spherical harmonics $Y_{lm}(\theta, \phi)$ and $f_n(\varphi)$
as the base functions. 
The ground-state eigenvalue and 
eigenfunction of Eq. (\ref{eq18}) can then be obtained numerically
for given applied force and torque
and the average extension  be calculated through 
Eq. (\ref{eq14}) or  through the following formula:
\begin{equation}
\langle Z\rangle =L \int \chi_0({\bf t}, \varphi) 
{\bf t}\cdot {\bf z}_0  \cos\varphi  \Phi_0({\bf t},\varphi) d {\bf t} d\varphi,
\label{supexten}
\end{equation}
where $\chi_0 ({\bf t}, \varphi)$ is the ground-state left-eigenfunction
of Eq. (\ref{eq18}).
\footnote{As 
remarked in Appendix \ref{appendixA}, because the operator  in
the square brackets of Eq. (\ref{eq18}) 
acting on $\Psi({\bf t}, \varphi; s)$
is not Hermitian, the resulting matrix form of the operator may 
not be diagonalized by unitary matrices. Consequently, in general
$\chi_0 ({\bf t},
\varphi)\neq \Phi_0^*({\bf t}, \varphi)$.}
The writhing number Eq. (\ref{eq10}) is calculated according to
Eq. (\ref{eqA16}) to be
\begin{equation}
\label{writhe2}
\langle Wr \rangle =L {\Gamma \over 16\pi \ell_p^* k_B T} 
\int \chi_0({\bf t}, \varphi)   \sin^2\theta \Phi_0({\bf t},\varphi) d{\bf t} 
d \varphi,
\end{equation}
and average  linking number is then calculated  to be
\begin{equation}
\label{link}
\langle Lk\rangle =\langle Tw\rangle+\langle Wr\rangle
={L \over 2\pi R} \int \chi_0 \sin\varphi \Phi_0 d{\bf t} d \varphi
+{L\Gamma\over 16\pi \ell_p^* k_B T} \int \chi_0 \sin^2\theta \Phi_0 d{\bf t}
d\varphi.
\end{equation}

Thus, after we have obtained
the ground-state eigenvalue as well as its left- and right-eigenfunction
numerically, we can calculate numerically all the quantities of our interest,
for example, the average extension, the average supercoiling degree, 
the folding angle distribution (see
also Appendix \ref{appendixA}).  
The relation between extension and supercoiling
degree can also be obtained by fixing the external force and changing the
value of the applied torque. 
To calculate the  ground-state eigenvalue and 
eigenfunctions of an asymmetric matrix turns out to be
complicated and time-consuming.  Fortunately, as we have
calculated  in Appendix {\ref{appendixB}}, each eigenfunction  of
Eq. (\ref{eq18}) shares the same quantity $m$; the matrix for $m=0$ is
still Hermitian and can be diagonalized by unitary matrices. The ground-state
eigenvalue for $m=0$ is  lower  in several order than those for
$m\neq 0$ in the whole relevant region of external  torque $\Gamma$ from
$-5.0 k_B T$ to $5.0 k_B T$. Thus, actually  we only  need to 
consider the case of $m=0$ and 
in this case we still have $\chi_0({\bf t},\varphi)=\Phi_0^*({\bf t},\varphi)$.
The whole procedure we have performed  in Sec. \ref{sec3} can safely  be
 repeated in this section, 
and the relationships  between force and extension and
between torque and linking number can be consequently 
calculated.

\begin{figure}[t]
\vskip 3.5cm
\hspace{1.2cm}\psfig{figure=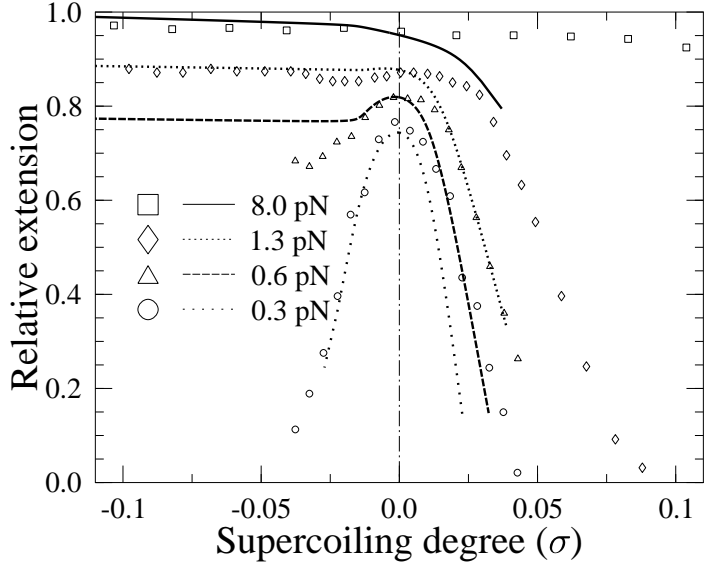,height=9.0cm}
\vskip -2.5cm
\caption{Extension vs   supercoiling relations
 at fixed pulling forces for torsionally constrained DNA.
The parameters for the  curves are the
same as  Fig.~{\protect\ref{fig2}} and experimental data is
from Fig.~3 of {\protect \cite{strick96}} (symbols).}
\label{fig5}
\end{figure}

To make the calculation further easier and also to
 make sure that the above-mentioned calculation is 
indeed correct, we introduce here an approximate method which  reduces the
computational complexity considerably. It turns out that  the calculated 
results using this method are in considerable agreement with the 
above-mentioned precise method.
In the experiment of Ref. \cite{strick96}, 
the applied external forces  change in the region 
of $0.3$ pN to $10$ pN. In this region, as 
demonstrated in Figs. \ref{fig3} and \ref{fig4}, 
both the  tangential (${\bf t}$) and the folding angle ($\varphi$)
  fluctuations  of dsDNA molecules are 
small. Taking into account this fact, then in Eq. (\ref{eq12}), the
energy term $\kappa^* (d{\bf t}/ds)^2$ can be approximately calculated to
be $\kappa^*((d t_x/ds)^2+(d t_y/ds)^2)$, and $f{\bf t}\cdot {\bf z}_0
\cos\varphi \simeq f\cos\varphi-f\langle\cos\varphi\rangle(t_x^2+t_y^2)/2$.
Thus, Eq. (\ref{eq12}) is decomposed into two $``$independent" parts.
The first part is only related to $\varphi$. At each value of
$f$ and $\Gamma$, we can calculate 
the average quantities $\langle \cos\varphi\rangle$ and
$\langle\sin\varphi\rangle$ based on this energy using
the method of path integral. 
The second part is   quadratic in $t_x$ and $t_y$, 
therefore the average values of 
${\bf t}\cdot {\bf z}_0$ and $Wr$ (Eq. (\ref{eq10}))
can be  obtained analytically. 
Using this decomposition and preaveraging technique, the average 
extension and average supercoiling degree   can both be calculated 
 at each value of external force and torque, and 
the relation between 
extension and linking number at
fixed forces can be then obtained.

The theoretical relationship between extension and
supercoiling degree is shown in Fig.\ \ref{fig5} and  compared
with the experiment of Strick {\it et al.}
\cite{strick96}.
In obtaining these curves, the values of the  parameters  
are the same as those used in Fig. \ref{fig2}
and  no adjustment has been done to fit the experimental data. 
We find that in the case of negatively supercoiled DNA, the
theoretical and experimental  results are
in quantitative agreement, indicating that the present model
is capable of explaining the elasticity of negatively supercoiled
DNA; in the case of positively supercoiled DNA, the agreement
between theory and experiment is not so good, especially
when the external force is relatively large.  
In our present work,
we have not considered the possible deformations of the nucleotide
basepairs. While this assumption might be reasonable in the
negatively supercoiled case, it may fail for positively supercoiled
DNA chain, especially at large   stretching  forces. 
The work done by Allemand {\it et al.} \cite{allemand98}
suggested that positive supercoiled and highly extended DNA molecule
can take on Pauling-like configurations with exposed bases. 
To better understand the elastic property of positively supercoiled 
DNA, it is certainly necessary for us to take into account the
deformations of basepairs.

For negatively supercoiled DNA molecule, 
both theory and experiment reveal the 
following elastic aspects: 
(a) When external force is small, 
DNA molecule can shake off its
torsional stress by writhing its 
central axis, which can lead to an increase in the 
negative writhing number and hence restore the local folding manner of
DNA strands to that of B-form DNA: 
(b) However, writhing of 
the central axis 
 causes shortening of DNA end-to-end extension, which becomes more and more
unfavorable as the external force is increased. Therefore, at large forces, 
the torsional stress caused by  negative torque (supercoiling degree) 
begins to unwind the B-form double-helix and triggers the transition
of DNA  internal structure, where  a continuously increasing portion of 
DNA takes on some certain new configuration as supercoiling increases,
while its total extension keeps almost invariant.   
Our Monte Carlo simulations have also confirmed the above insight 
\cite{zhangy99}.

\begin{figure}[t]
\vskip 3.5cm
\hspace{1.2cm}\psfig{figure=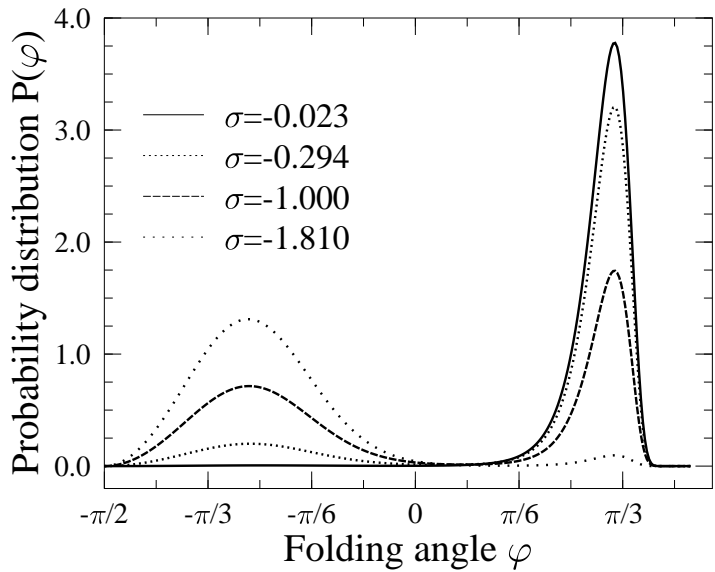,height=9.0cm}
\vskip -2.5cm
\caption{Folding angle distributions for negatively supercoiled
 DNA molecule pulled with a force of \protect\(1.3\protect\) pN.}
\label{fig6}
\end{figure}

What is the new configuration?
According to the opinion of Ref. \cite{strick96}, such new configuration
corresponds to denatured DNA segments, i.e., negative torque leads to
breakage  of hydrogen bonds between the complementary DNA bases and 
consequently to strand-separation. They have also done an elegant 
experiment in which short single-stranded homologous DNA segments are
inserted into the experimental buffer \cite{strick98}. They found that,
in confirmative with their insight, these homologous DNA probes indeed
bind onto negatively supercoiled dsDNA molecules. Recently,
L${\rm\acute{e}}$ger {\it et al.} \cite{leger99} 
have also done experiment on single dsDNA molecules. 
To explain qualitatively their experimental result, they
 found that left-handed Z-form DNA should be considered as
a possible configuration for negatively supercoiled dsDNA chain, 
while the molecules need not be denatured. 
As seen in Fig. \ref{fig5}, although the present model has
not taken into account the possibility of strand-separation, it
can quantitatively explain the behavior of negatively supercoiled
DNA. Therefore, it may be helpful for us to 
  investigate the possibility of 
formation of left-handed configurations based on our
present model. 
This effort is done in the next subsection, where  the energetics of 
such configurations will also be discussed. 

\subsection{Possible left-handed DNA configurations}
\label{sec4B}  

We have mentioned in Sec. \ref{sec2A} that left-handed 
configurations correspond to $\varphi <0$ in our present
model (see also Fig. \ref{fig1}). Based on the present model, then 
information about the  new configuration mentioned in Sec.
\ref{sec4A} can be revealed by
the folding angle   distributions $P(\varphi)$, as have been
discussed in Sec. \ref{sec3}.
In the present case, $P(\varphi)$ is calculated as follows:
\begin{equation}
\label{folding2}
P(\varphi)=\int \chi_0({\bf t},\varphi)\Phi_0({\bf t},\varphi) d{\bf t},
\end{equation}
where, as mentioned in Sec. \ref{sec4A}, 
$\Phi_0({\bf t}, \varphi)$ and $\chi_0({\bf t},\varphi)$ are,
respectively, the ground-state right- and left-eigenfunction of
Eq. (\ref{eq18}); and actually, $\chi_0({\bf t},\varphi)=
\Phi_0^*({\bf t},\varphi)$.

The calculated folding angle distribution \cite{jcp} is shown in
Fig. \ref{fig6}, which is radically  different with 
 that of torsionally relaxed dsDNA molecules
shown in Fig. \ref{fig3}. It has the following aspects:
When the torsional stress is small (with the supercoiling degree
$|\sigma| < 0.025$), the distribution has only one narrow and steep peak
at $\varphi \simeq + 57.0^\circ$, indicating that  DNA is
completely in  B-form.
 With the increase of  torsional 
stress, however, another peak appears at $\varphi \simeq
- 48.6^\circ$ and the total probability for the folding angle to be 
negatively-valued increases   gradually with supercoiling.
Since negative folding angles correspond to left-handed configurations,
the present model suggests that, with the increasing of supercoiling, 
left-handed DNA conformation is nucleated and it
then elongates along the DNA chain
as B-DNA disappears gradually. The whole chain becomes completely
left-handed at $\sigma\simeq -1.85$.

\begin{figure}[t]
\vskip 3.5cm
\hspace{-0.5cm}\psfig{figure=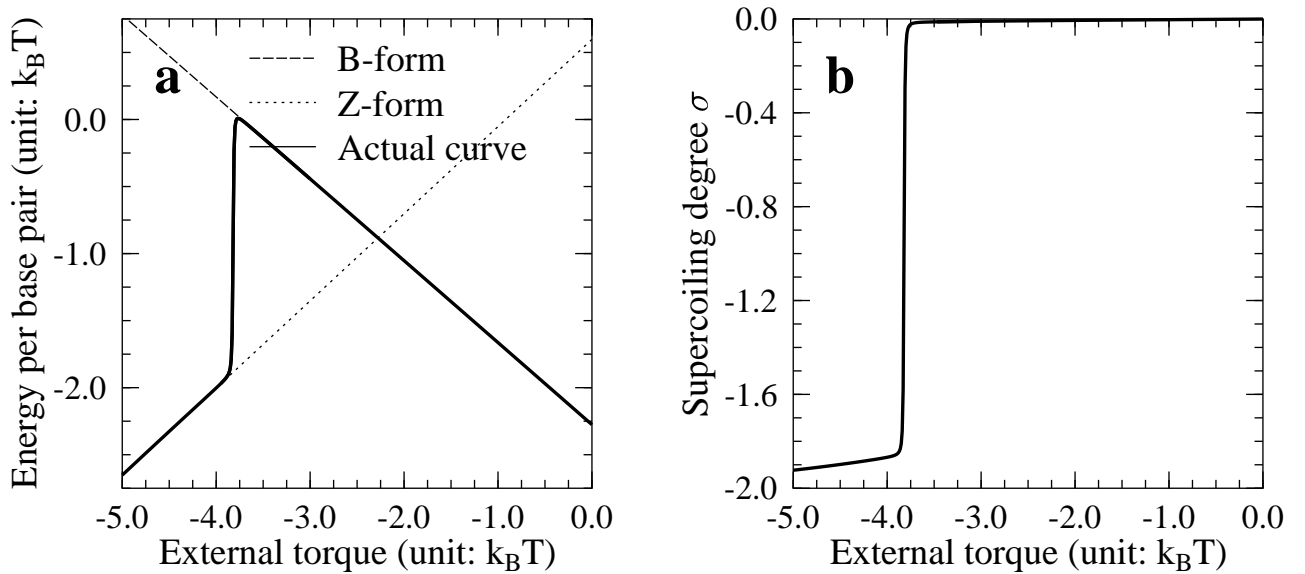,height=9.0cm}
\vskip -3.5cm
\caption{ (A),
The sum of the average base-stacking and torsional energy per
basepair  at force \protect\(1.3\protect\) pN.
For highly extended DNA only these two  interactions
are sensitive with torque.
(B), The relation between DNA supercoiling degree and external
torque at force  \protect\(1.3\protect\) pN.}
\label{fig7}
\end{figure}

It is worth to be noticed  that, (a) as the supercoiling degree 
changes, the positions of the two peaks of the folding angle distribution
remain almost fixed and, 
(b) between these two peaks, there exists an extended region
of folding angle from $0$ to $\pi/6$ which always has only extremely small
probability of occurrence. Thus, a negatively  supercoiled DNA 
can have two possible stable configurations, a right-handed B-form  
and a left-handed configuration with an average 
folding angle $\simeq -48.6^\circ$.
A transition between these two structures for a DNA segment
will generally lead to an abrupt and finite variation in the folding angle.

To obtain the energetics of such transitions, we have
calculated  how the 
sum of the  base-stacking energy and torsional energy,
$(\kappa/R^2)\sin^4\varphi+\rho(\varphi)-(\Gamma/R)\sin\varphi$,
 changes with external torque \cite{jcp}. Figure \ref{fig7}a 
shows the numerical result, and Fig. \ref{fig7}b 
demonstrates the relation between supercoiling degree and 
external torque. (In both figures the external force is
fixed at $1.3$ pN.) 
 From these figures we can infer that, (a)
 for negative torque  less than the
critical value $\Gamma_c\simeq -3.8$ 
k$_B$T, DNA can only stay in B-form state; 
(b) near this critical torque,  DNA can either
be right- or be left-handed and, as negative
supercoiling increases (see Fig. \ref{fig7}b) more and more DNA segments
will stay in the left-handed form, which is much lower in
energy ($\simeq -2.0$ k$_B$T per basepair)
 but stable only  when torque reaches $\Gamma_c$;
(c) for negative torque greater than $\Gamma_c$ DNA is    
completely left-handed.

Nevertheless, we should emphasize that the above 
calculations are all based on our present model which 
has assumed that nucleotide basepairs do not break.
 Figure \ref{fig5} indicates that for 
negatively supercoiled DNA chains, the extension vs supercoiling degree
relation can be quantitatively explained by the present model; 
and Fig. \ref{fig6} reveals the reason of the quantitative 
agreement is that the present 
model allows the possibility of occurrence 
of left-handed DNA configurations.  
At the present time, 
to say that negatively supercoiled DNA will prefer
left-handed configurations rather than  denaturation and 
strand-separation  is 
premature. To clarify this question, 
the present model should be improved to
consider the deformations of DNA basepairs.
In the experimental side, it might also be helpful to 
measure precisely the critical torque at which the elastic
behavior of negatively supercoiled DNA changes abruptly and
compare the measured results with  the value calculated in
the present work. (In the earlier experiment of Allemand {\it et al.}
\cite{allemand98}, the critical torque is estimated to be
$\sim -2 k_B T$ by assuming that the torsional rigidity of dsDNA to be
$75$ nm and that torsional stress builds up linearly along the
DNA chain.  This value, however, maybe not precise enough since
the torsional rigidity of dsDNA molecule is not a precisely
determined quantity and the values giving by  different groups   
are scattered widely.\footnote{Another possibility may be that we have
over-estimated the value of the critical torque.
 It maybe possible that the transition from right- to left-handed
configurations is initiated in  the weaker AT-rich regions, whose 
value of $\epsilon$ should be less than the average value  taken by the
present paper.}

The structural parameters of the   
 left-handed configuration suggested by  Figs. \ref{fig6} and \ref{fig7} 
are listed in table \ref{table1} \cite{jcp} and compared with
those of Z-form DNA \cite{watson87}. The strong similarity
in these parameters suggests that the torque-induced 
left-handed configurations, if they really exist,
belong to Z-form DNA \cite{watson87,leger99,jcp}.

\begin{table}[t]
\vskip 1.0cm
\footnotesize
\begin{center}
\caption{For the torque-induced left-handed
DNA configuration, the average rise per basepair ($d$),
the pitch per turn of helix, and the number of basepairs
per turn of helix ($num$) are calculated and listed  under
different external forces and torques. The last row
contains the corresponding values for Z-form DNA. }
\vskip 0.5cm
\begin{tabular}{|ccccc|}
\label{table1}
force (pN) & torque ($k_B T$) & $d$ (\AA) & pitch (\AA) & $num$ \\
\hline
1.3 & -5.0 & 3.59 & 41.20 & 11.48 \\
1.0  & -5.0 & 3.57 & 40.93 & 11.44 \\
1.3 & -4.0 & 3.83 & 46.76 & 12.19 \\
1.0 & -4.0 & 3.82 & 46.38 & 12.15 \\
\hline
Z-form: & & 3.8 & 45.6 & 12
\end{tabular}
\end{center}
\end{table}

\section{conclusion}
\label{sec5}

In this article, we have presented an elastic model for
double-stranded biopolymers such as DNA molecules. The
key progress is that the bending 
deformations of the backbones of DNA molecules and the 
base-stacking interactions existing between adjacent DNA basepairs
are quantitatively considered in this model, with the introduction
of a new structural parameter, the folding angle $\varphi$. 
This model has also qualitatively taken into account the effects of
the steric effects of DNA basepairs and electrostatic interactions 
along DNA chain. 
In calculation technique, the model is investigated using path integral 
method; and Green equations similar in form to the Schr${\rm\ddot{o}}$dinger
equation in quantum mechanics
 are derived and their ground-state eigenvalues
and eigenfunctions obtained by precise numerical calculations.
The force-extension relationship in torsionally relaxed and the
extension-linking number relationship in torsionally constrained
DNA chains are studied and compared with experimental results. 
This work demonstrated that DNA molecule's entropic elasticity and
 highly extensibility, as well as the elastic property of
negatively supercoiled DNA can all be quantitatively explained 
by the present theory. The comprehensive agreement
between theory and experiments indicated that the short-ranged
base-stacking interactions are very important in determining the
elastic response of double-stranded DNA molecules. The
present work showed that highly extended and negatively supercoiled DNA
molecules can be left-handed, probably in the Z-form configuration.
A possible way to check the validity of this opinion is to
measure the critical external torque at which the transition
between  B-form DNA and the new configuration takes place.

The present work regarded DNA basepairs as rigid objects and
did  not consider their possible deformations and the 
possibility of strand-separation.  
The comprehensive agreement between theory and experiments 
indicates that  this approach is well justified in
many cases.  However,
as already mentioned in Sec. \ref{sec2A}, under some extreme conditions,
this assumption may  not be  appropriate. For example, recent 
experimental work of Allemand {\it et al.} \cite{allemand98} showed
that positively supercoiled DNA under high applied force
can take on Pauling-like configurations
with exposed bases. In this case the basepairing of DNA is 
severely distorted, and because the present model has not taken
into account the possible deformations of the basepairs, the theoretical
results on positively supercoiled DNA molecules were
not in quantitative agreement with experiment 
(see Fig. \ref{fig5}). Furthermore, although
the present work showed that left-handed 
DNA configurations can be stabilized by negative torques,
much theoretical work is still needed to 
 calculate the denaturation free energy and
be compared with the free energy of left-handed DNA configurations.

\section{Acknowledgement}
\label{acknow}
It gives us great pleasure to acknowledge the helps and valuable
suggestions of our  colleagues, especially Lou Ji-Zhong, Liu Lian-Shou,
Yan Jie, Liu Quan-Hui, Zhou Xin, Zhou Jian-Jun, and Zhang Yong. 
The numerical calculations are performed partly at ITP-Net and
partly   at the State Key Lab. of
Scientific and Engineering Computing.

\appendix
\section{Path integral method in polymer physics}
\label{appendixA}

In this appendix we review some basic ideas on the application of 
path integral method to the study of polymeric systems \cite{zhounote1}.
Consider a polymeric string, and suppose its total $``$arclength" is $L$, and
along each arclength point $s$ one can define a n-dimensional $``$vector"
${\bf r}(s)$ to describe the polymer's local state at this 
point.\footnote{For example, in the case of a flexible Gaussian chain,
${\bf r}$ is a three-dimensional position vector; in the
case of a semiflexible chain  such as the wormlike chain
\cite{marko95a,kroy96}, 
${\bf r}$ is the unit tangent vector of the polymer and is
two-dimensional.}
We further assume that the energy density (per unit arclength) of the
polymer can be written as the following general form:
\begin{equation}
\label{eqA1}
\rho_e({\bf r},s)={m \over 2}({d{\bf r}\over ds})^2+{\bf A}({\bf r})\cdot
{d{\bf r}\over ds}+V({\bf r}),
\end{equation}
where $V({\bf r})$ is a scalar field and ${\bf A}({\bf r})$ is a
vectorial field.  
The total partition function of the system is expressed by 
the following integration:
\begin{equation}
\label{eqA2}
\Xi(L)=\int\int d {\bf r}_f \phi_f({\bf r}_f) G({\bf r}_f, L; {\bf r}_i,
0) \phi_i ({\bf r}_i) d {\bf r}_i,
\end{equation}
where $\phi_i({\bf r})$ and $\phi_f({\bf r})$ are, respectively, the
probability distributions of the vector ${\bf r}$ at the initial ($s=0$)
and final ($s=L$) arclength point; $G({\bf r}, s; {\bf r}^\prime, s^\prime)$
is called the Green function, it is 
defined in the following way:
\begin{equation}
\label{eqA3}
G({\bf r}, s; {\bf r}^\prime, s^\prime)
=\int_{{\bf r}^\prime}^{{\bf r}} {\cal D}[{\bf r}^{\prime\prime}(s)]
\exp[-\beta \int_{s^\prime}^s d s^{\prime\prime} \rho_e({\bf r}^{\prime\prime},
s^{\prime\prime}) ],
\end{equation}
where integration is carried over all possible configurations of
${\bf r}^{\prime\prime}$, and $\beta=1/k_B T$ is the
Boltzmann coefficient.
It can be  verified that the Green function defined above satisfies
the following relation \cite{zhounote1}:
\begin{equation}
\label{eqA4}
G({\bf r}, s; {\bf r}^\prime, s^\prime)=
\int d {\bf r}^{\prime\prime} G({\bf r}, s; {\bf r}^{\prime\prime}, 
s^{\prime\prime}) G({\bf r}^{\prime\prime}, s^{\prime\prime}; 
{\bf r}^\prime, s^\prime), \;\;\;\;\; (s^\prime < s^{\prime\prime} < s).
\end{equation}
The total free energy of the system is then expressed as
\begin{equation}
\label{eqA5}
{\cal F}=-k_B T \ln \Xi.
\end{equation}

To calculate the total partition function $\Xi$, we define an 
auxiliary function $\Psi({\bf r}, s)$ and call it the wave function because
of its similarity with the true wave function of quantum systems. 
Suppose the value of $\Psi$ at 
arclength point $s$ is related to its value at $s^\prime$ through
the following formula such  
that\footnote{In fact, the choice of the wave function $\Psi({\bf r}, s)$
is not limited. Any function  
determined by an integration of the form of Eq. (\ref{eqA6}) can
be viewed as a wave function.}
\begin{equation}
\label{eqA6}
\Psi({\bf r}, s)=\int d {\bf r}^\prime G({\bf r}, s; {\bf r}^\prime,
s^{\prime}) \Psi({\bf r}^\prime, s^\prime), \;\;\;\;\; (s> s^\prime)
\end{equation}
then, we  can derive from Eqs. (\ref{eqA1}), (\ref{eqA3}), and
(\ref{eqA6}) that \cite{zhounote1}:
\begin{equation}
\label{eqA7}
{\partial \Psi({\bf r}, s)\over \partial s}=
\left[{{\bf \nabla}_{\bf r}^2 \over 2 m \beta}-\beta V({\bf r})+
{{\bf A}({\bf r}) \cdot {\bf \nabla}_{\bf r} \over m}+
{{\bf \nabla}_{\bf r} \cdot {\bf A}({\bf r}) \over 2m}
+ {\beta {\bf A}^2({\bf r}) \over 2 m}\right] \Psi({\bf r}, s)
=\hat{H} \Psi({\bf r}, s).
\end{equation}
Equation  (\ref{eqA7}) is called the Green equation,
it is very similar with the  Schr${\rm\ddot{o}}$dinger equation 
of  quantum mechanics \cite{zeng99}.  
However, there is an important difference.  
In the case of ${\bf A}({\bf r}) \neq 0$, the operator $\hat{H}$
in Eq. (\ref{eqA7}) is not Hermitian.  Therefore, in this 
case the matrix form of the operator $\hat{H}$ may not be
diagonalized   by unitary matrix. 

Denote the eigenvalues and the right-eigenfunctions of 
Eq. (\ref{eqA7}) as $-g_i$ and
$|i \rangle=\Phi_i ({\bf r})$ ($i=0, 1, \cdots$), respectively. 
Then it is easy to know, from the
approach of quantum mechanics,  that 
\begin{equation}
\label{eqA8}
\Psi(s)=\sum\limits_{i} e^{-g_i (s-s^\prime)} |i(s)\rangle\langle
i(s^\prime)|\Psi(s^\prime)\rangle,
\end{equation}
where $\langle i|=\chi_i({\bf r})$ $(i=0, 1,\cdots)$ 
denote the left-eigenfunctions of Eq. (\ref{eqA7}),
which satisfy  the following relation:
$$
\langle i| i^\prime \rangle=\int d{\bf r} \chi_i ({\bf r})\Phi_{i^\prime}
({\bf r}) =\delta_i^{i^\prime}.
$$ 
In the case where 
$\hat{H}$ is Hermitian (i.e., ${\bf A}({\bf r})=0$), then we can
conclude that 
$$
\chi_i({\bf r})=\Phi_i^*({\bf r}).
$$

From Eqs. (\ref{eqA2}), (\ref{eqA6}), (\ref{eqA7}), 
and (\ref{eqA8}) we know that
\begin{eqnarray}
\label{eqA9}
 \Xi(L) & = & \int d{\bf r} 
\int d{\bf r}^\prime G({\bf r}, L; {\bf r}^\prime, 
s^\prime) \phi_f({\bf r}) \phi_i ({\bf r}^\prime) = 
\sum\limits_{i} \langle \phi_f|i\rangle\langle i|\phi_i\rangle
e^{-g_i L} 
\nonumber \\
 & = &  e^{-g_0 L} \langle \phi_f|0\rangle\langle 0|\phi_i\rangle \;\;\;\;
\;\;({\rm for}\;\; L\gg 1/(g_1-g_0) ).
\end{eqnarray}
Consequently, for long polymer chains the total free energy density is just
expressed as
\begin{equation}
{\cal F}/L= k_B T g_0,
\label{eqA10}
\end{equation} 
and any quantity of interest can  then be  calculated by
differentiation of ${\cal F}$. For example, the average  extension 
of a polymer under external force field $f$ can be calculated as
$\langle Z \rangle=
\partial {\cal F}/\partial f = L k_B T \partial g_0/\partial f$.

We continue to discuss another very important quantity,
the  distribution probability 
 of ${\bf r}$ at arclength $s$, $P({\bf r}, s)$.
 This probability is calculated from
the following expression:
\begin{equation}
\label{eqA11}
P({\bf r}, s)={
{ \int d{\bf r}_f \int d {\bf r}_i \phi_f({\bf r}_f)G({\bf r}_f, L;
{\bf r}, s) G({\bf r}, s; {\bf r}_i, 0) \phi_i({\bf r}_i)}
\over 
{\int d {\bf r}_f \int d {\bf r}_i \phi_f({\bf r}_f) G({\bf r}_f, L;
{\bf r}_i, 0) \phi_i({\bf r}_i)}
}.
\end{equation}
Based on Eqs. (\ref{eqA6}) 
and (\ref{eqA8}) we can rewrite Eq. (\ref{eqA11})
in the following form:
\begin{eqnarray}
\label{eqA12}
P({\bf r}, s) & = & {{
\int d{\bf r}_f \int d{\bf r}_i \int d {\bf r}^\prime
\phi_f({\bf r}_f) G({\bf r}_f, L; {\bf r}^\prime, s)\delta ({\bf r}^\prime
-{\bf r}) G({\bf r}, s; {\bf r}_i, 0) \phi_i ({\bf r}_i)}
\over {\int d {\bf r}_f \int d {\bf r}_i \phi_f({\bf r}_f) G({\bf r}_f, L;
{\bf r}_i, 0) \phi_i({\bf r}_i)}
} \nonumber \\
 & = & {
 {\sum\limits_m \sum\limits_n \langle \phi_f | m\rangle \langle n |\phi_i
\rangle \chi_m({\bf r})\Phi_n({\bf r}) \exp[-g_m(L-s)-g_n s]}
\over {\sum\limits_m \langle \phi_f| m\rangle \langle m|
\phi_i\rangle \exp(-g_m L)}}
\end{eqnarray}
For the most important case of $0\ll s\ll L$, Eq. (\ref{eqA12}) then
gives that the probability distribution of ${\bf r}$ is independent
of arclength $s$, i.e.,
\begin{equation}
\label{eqA13}
P({\bf r}, s)=\chi_0({\bf r})\Phi_0({\bf r}) \;\;\;\;\;({\rm for}\;\;
0\ll s \ll L).
\end{equation}
 With the help of Eq. ({\ref{eqA13}), the average value of 
a quantity which is a function of ${\bf r}$ can be obtained. For example,
\begin{equation}
\label{eqA14}
\langle Q(s)\rangle =\int d {\bf r} Q({\bf r}) P({\bf r}, s)  
=\int d{\bf r} \chi_0({\bf r})Q({\bf r})\Phi_0({\bf r})
=\langle 0|Q|0\rangle,
\end{equation}
and
\begin{equation}
\label{eqA15}
\langle \int_0^L Q({\bf r}(s)) d s\rangle
=\int_0^L \langle Q(s)\rangle ds 
=L \langle 0|Q|0\rangle \;\;\;\;({\rm for}\;\;
L\gg 1/(g_1-g_0)).
\end{equation}

Finally, we list the formula for  calculating $\langle {\bf B}({\bf r})
\cdot d{\bf r}/ds\rangle$, here ${\bf B}({\bf r})$ is a given vectorial field.
The formula reads:
\begin{equation}
\label{eqA16}
\langle {\bf B}({\bf r})\cdot {d{\bf r}\over ds}\rangle
={1\over 2} \langle 0 | \left[ {\bf r}\cdot (\hat{H} {\bf B})
-{\bf B}\cdot (\hat{H} {\bf r}) \right] | 0 \rangle
\;\;\;\; ({\rm for}\;\; L\gg 1/(g_1-g_0)).
\end{equation}

\section{Matrix formalism of the Green Equation Eq. (23) }
\label{appendixB}

In Eq. (\ref{eq18}) (and also Eq. (\ref{eq13})) the
variables ${\bf t}$ and $\varphi$  are coupled together. 
Denote the operator in the square brackets of Eq. (\ref{eq18}) as 
$\hat{H}$.  We express this operator in matrix form. To this
end we choose the base functions of this system to be the
combinations of spherical harmonics $Y_{lm}(\theta,\phi)$ and 
$f_n(\tilde{\varphi})$ (see Eq. (\ref{fn})).  

For $m=0$, the  base functions are expressed to be
\begin{equation}
\label{eqB1}
|i\rangle =|l\cdot N_{\varphi}+n\rangle =
|l;n\rangle = Y_{l 0}(\theta,\phi) f_n(\tilde{\varphi});
\end{equation}
and for $m=1,2,\cdots$, the base functions are expressed to be
\begin{equation}
\label{eqB2}
|i\rangle =|2 (l-m)\cdot N_{\varphi}+2 (n-1)+k \rangle
=|l;n;k\rangle =Y_{lm}^{(k)}(\theta,\phi)f_n(\tilde{\varphi}).
\end{equation}
In the above two equations, $n=1,2,\cdots, N_{\varphi}$ with
$N_{\varphi}$ set to be $60$ in our present calculations; $l=0,1,\cdots,
N_l -1$ for the case of $m=0$ (with $N_l$ set to $30$), or $l=m,m+1,
\cdots,m+N_l-1$ for the case of $m\neq 0$ (with $N_l$ set to $15$); 
$k=1,2$; and
\begin{eqnarray}
Y_{l0}(\theta,\phi) & = & \sqrt{{2l+1\over 4\pi}}P_l(\cos\theta), \nonumber \\
Y_{lm}^{(1)}(\theta,\phi) & = & (-1)^m 
\sqrt{{2 l+1 \over 2 \pi}{(l-m)!\over (l+m)!}}P_l^m(\cos\theta)\cos(m\phi), \nonumber \\
Y_{lm}^{(2)}(\theta,\phi) & = & (-1)^m
\sqrt{{2l+1\over 2\pi}{(l-m)!\over (l+m)!}} P_l^m(\cos\theta)\sin(m\phi). \nonumber
\end{eqnarray}

With the above definitions, we then obtain
that, for $m=0$:
\begin{eqnarray}
\langle i_p | (-\hat{H})| i\rangle &  =  & \langle l_p;n_p| (-\hat{H})
| l; n\rangle \nonumber \\
  & = & \delta_l^{l_p}\delta_n^{n_p} \left[{l(l+1)\over 4\ell_p^*}
+{n^2 \pi^2\over 4\ell_p a^2} \right]  \nonumber \\
 & - & {f\over k_B T} \left[\delta_{l_p}^{l+1} a_{l,0}+
\delta_{l_p}^{l-1} a_{l-1,0}\right]\langle n_p|\sin\tilde{\varphi}|n\rangle \nonumber \\
 & + & \delta_{l_p}^l \langle n_p |\left[{\rho(\tilde{\varphi})\over k_B T}
+{\ell_p \over R^2}\cos^4\tilde{\varphi}\right] | n\rangle \nonumber \\
 & + & \delta_{l_p}^l {\Gamma \over R k_B T}
\langle n_p | \cos\tilde{\varphi} | n \rangle \nonumber \\
 & - & \delta_{n_p}^n {\Gamma^2 \over 16 \ell_p^* (k_B T)^2} 
\left[\delta_{l_p}^l (1-a_{l,0}^2-a_{l-1,0}^2) 
-\delta_{l_p}^{l+2} a_{l,0} a_{l+1,0} -\delta_{l_p}^{l-2} a_{l-1,0} a_{l-2,0}\right];
\label{eqB3}
\end{eqnarray}
and for $m\neq 0$:
\begin{eqnarray}
\langle i_p | (-\hat{H})| i\rangle &  =  & \langle l_p;n_p;k_p| (-\hat{H})
| l; n;k \rangle \nonumber \\
  & = & \delta_{l_p}^{l}\delta_{n_p}^{n}\delta_{k_p}^k
 \left[{l(l+1)\over 4\ell_p^*}
+{n^2 \pi^2\over 4\ell_p a^2} \right]  \nonumber \\
 & - & {f\over k_B T}\delta_{k_p}^k \left[\delta_{l_p}^{l+1} a_{l,m}+
\delta_{l_p}^{l-1} a_{l-1,m}
\right]\langle n_p|\sin\tilde{\varphi}|n\rangle \nonumber \\
 & + & \delta_{l_p}^l \delta_{k_p}^k
 \langle n_p |\left[{\rho(\tilde{\varphi})\over k_B T}
+{\ell_p \over R^2}\cos^4\tilde{\varphi}\right] | n\rangle \nonumber \\
 & + & \delta_{l_p}^l\delta_{k_p}^k {\Gamma \over R k_B T}
\langle n_p | \cos\tilde{\varphi} | n \rangle 
+\delta_{l_p}^l \delta_{n_p}^n {\Gamma \over
4 \ell_p^* k_B T} (k-k_p) m \nonumber \\
 & - & \delta_{n_p}^n \delta_{k_p}^k 
{\Gamma^2 \over 16 \ell_p^* (k_B T)^2} 
\left[\delta_{l_p}^l (1-a_{l,m}^2-a_{l-1,m}^2) 
-\delta_{l_p}^{l+2} a_{l,m} a_{l+1,m} -\delta_{l_p}^{l-2}
 a_{l-1,m} a_{l-2,m}\right].
\label{eqB4}
\end{eqnarray}
In Eqs. (\ref{eqB3}) and (\ref{eqB4}),  the following notations are
used:
\begin{eqnarray}
 & & a_{l,m}=\sqrt{{(l+1)^2-m^2 \over (2 l+1) (2 l+3)}}, \nonumber \\
 & & \langle n_p | y(\tilde{\varphi}) | n \rangle =
{2 \over a} \int_0^a \sin({n_p \pi \tilde{\varphi} \over a}) y(\tilde{\varphi})
\sin({n\pi \tilde{\varphi} \over a}), \nonumber
\end{eqnarray}
where $y(\tilde{\varphi})$ is any function of $\tilde{\varphi}$. 

The ground-state eigenvalues of the matrices Eq. (\ref{eqB3}) and 
Eq. (\ref{eqB4}) have been  calculated at force $1.3$ pN
 in the whole relevant region of 
$\Gamma$ from $-5.0 k_B T $ to $+5.0 k_B T$. 
We have found that the ground-state eigenvalues for  the case
of $m=0$ are  of the order of $-10$ nm$^{-1}$; while those  for
the case of $m\neq 0$  are of the order of $10^{6}$ nm$^{-1}$ 
(data not shown).   Thus, we can conclude with confidence that
the $``$low energy" eigenstates of the system all have the 
same $m=0$. This can greatly reduce the calculation tasks.

As a final point, here we demonstrate how to calculate the  average values of 
the quantities of our interest. Suppose $y({\bf t},\tilde{\varphi} )$ 
is a quantity  whose average value we are interested in.
Denote ${\cal U}$ as the unitary matrix which can diagonalize
the matrix Eq. (\ref{eqB3}). (Because 
the matrix Eq. (\ref{eqB3}) is real symmetric,
${\cal U}$ is actually a real orthogonal matrix.)
 Then its column vector ${\bf u}(i) =
 \sum_{j}{\cal U}(j,i) |j\rangle$  $(i=1,2,\cdots)$ corresponds to  
the $i$-th eigenvector of Eq. (\ref{eqB3}). Consequently,
we can calculate based on Eq. (\ref{eqA14}) that 
\begin{equation}
\langle y({\bf t},\tilde{\varphi})\rangle =
\sum\limits_{i,i_p} {\cal U}(i_p,1) {\cal U}(i,1)\langle i_p | y({\bf t},\tilde{\varphi})
| i\rangle.
\end{equation}

\end{document}